# On the limitations of probabilistic claims about the probative value of mixed DNA profile evidence

**10 Sept 2020**

Norman Fenton[1,3], Allan Jamieson[2], Sara Gomes[2], Martin Neil[1,3]

[1] Risk and Information Management, Queen Mary University of London, United Kingdom
[2] The Forensic Institute
[3] Agena Ltd, Cambridge, UK

## Abstract

The likelihood ratio (LR) is a commonly used measure for determining the strength of forensic match evidence. When a forensic expert determines a high LR for DNA found at a crime scene matching the DNA profile of a suspect they typically report that 'this provides strong support for the prosecution hypothesis that the DNA comes from the suspect'. However, even with a high LR, the evidence might ***not support*** the prosecution hypothesis if the defence hypothesis used to determine the LR is not the negation of the prosecution hypothesis (such as when the alternative is 'DNA comes from a person unrelated to the defendant' instead of 'DNA does not come from the suspect'). For DNA mixture profiles, especially low template DNA (LTDNA), the value of a high LR for a 'match' – typically computed from probabilistic genotyping software - can be especially questionable. But this is not just because of the use of non-exhaustive hypotheses in such cases. In contrast to single profile DNA 'matches', where the only residual uncertainty is whether a person other than the suspect has the same matching DNA profile, it is possible for all the genotypes of the suspect's DNA profile to appear at each locus of a DNA mixture, even though none of the contributors has that DNA profile. In fact, in the absence of other evidence, we show it is possible to have a very high LR for the hypothesis 'suspect is included in the mixture' even though the posterior probability that the suspect is included is very low. Yet, in such cases a forensic expert will generally still report a high LR as 'strong support for the suspect being a contributor'. Our observations suggest that, in certain circumstances, the use of the LR may have led lawyers and jurors into grossly overestimating the probative value of a LTDNA mixed profile 'match'.

Corresponding Author: Norman Fenton n.fenton@qmul.ac.uk

# 1. Background

The observation that DNA profiles could be obtained from many everyday objects [1], and the development of techniques to enable Short Tandem Repeat (STR) profiles from a single cell [2] were important advances in forensic science. Many forensic samples comprise cellular material from more than one person. Guidelines for the interpretation of two-person mixtures were published in [3]; this method excluded some genotypes on the basis that the peak heights made some genotype combinations impossible, or at least highly unlikely. . The analysis of 'mixed profiles' has been, and is, accomplished statistically in different ways. This paper deals with only one of those; the Likelihood Ratio (LR). There has been a plethora of publications promoting the use of the LR as one of the methods of evaluating forensic scientific evidence [4]–[6] and some promoting it as the only one [7] .
The LR approach is discussed herein, for clarity, using primarily examples with unambiguous allele designations, although many casework samples may be wholly or partially comprised of uncertain alleles.

DNA may be compromised in various ways thereby reducing the amount of information that can be obtained; for example, inhibition, degradation or insufficient amount of DNA. Such compromised profiles generally include very small peaks at one or more loci on the electropherogram (epg), which is the graph that DNA analysts use to decide which components (alleles) are present in a sample. While all types of DNA analyses can be compromised by problems of contamination, secondary transfer and generally poor handling procedures, there are special problems associated with samples where one or more of the contributors' profiles are subject to these effects. After some initial confusion between the technique used to create the profile and the method used to interpret the profile, it is now widely accepted that samples requiring different interpretation guidelines because of low amounts of DNA from one or more contributors in the sample are described as Low Template DNA (LTDNA) [8]; also, "The fundamental difference between DNA analysis of complex-mixture samples and DNA analysis of single-source and simple mixtures lies not in the laboratory processing, but in the interpretation of the resulting DNA profile." [9].

LTDNA analyses are characterised by increased 'stochastic effects' which are caused by sampling from the smaller number of DNA molecules present in the sample. Alleles may be wrongly recorded as missing (called "dropout") or wrongly recorded as present (called "dropin") with highly variable allelic and stutter peak heights (stutters are artefacts which can be wrongly identified as alleles. While it is possible (and usual) to apply replicate analyses to address these concerns, replication divides an already tiny sample into even smaller aliquots, and typically produces different results each time [10]. In reporting their findings forensic scientists are not obliged to provide all results or to attempt to combine them.

In mixed samples, it is impossible to determine conclusively the number of people in the mixture. Although it has been claimed that misassignment of the number of contributors to a profile has no significant effect [11], in practical casework conclusions about whether a particular person is or is not likely to be a contributor can drastically change if the assumed number of contributors is changed. For example, we have had instances in casework where, if the defendant is proposed as a contributor, then an additional contributor must be added to the prosecution hypothesis.

Hence, unsurprisingly therefore, the reliability of LTDNA evidence has been questioned in court and a number of high-profile cases that rely on such evidence have collapsed [12], [13] To address these concerns, and to deal with the inevitable subjectivity and uncertainty associated with interpreting LTDNA profiles, it is now becoming routine to treat allele detection as a continuous (using peak height) rather than a binary (also called semi-continuous or discrete; present/absent) identification problem and to use probabilistic methods to determine the overall probative value of the evidence [12], [14]–[16]. These techniques have largely been automated in probabilistic genotyping software tools such as [17]–[21] and in most approaches the likelihood ratio (LR) is the preferred method for quantifying probative value. The probabilistic approach may address some of the above concerns, and is especially useful in situations where:

- It is clear that there is a known contributor (such as the victim) and a single unknown contributor.
- The 'size' of each different unknown contributor's profiles are very different (for example, approximately 70%, 25% and 5% respectively for three contributors)

However, these are also situations where analysts have for decades been able to interpret such profiles without the need for additional systems [3]. The ability to resolve such profiles does not imply the ability to resolve more complex profiles. Concerns about the reliability of probabilistic genotyping methods and software have been raised [22], [23] including the difficulty of validating complex proprietary software and the fact that different tools produce different results for the same sample [24], [25], and the same programme producing different results from the same sample [26]. In this paper we will focus on higher level concerns about the use and interpretation of the LR that are independent of whether binary or continuous (probabilistic) allele identification is used, or indeed any of the numerous software 'solutions'. Specifically, the focus will be on circumstances for mixed profiles under which the LR information may be misleading and the conclusions typically drawn from it may be wrong. The implications of this problem are not widely understood by forensic scientists or members of the judiciary.

The paper is structured as follows: In Section 2 we provide an overview of the LR and the general problems associated with its use and interpretation, including the limitations of the LR in cases where the paired hypotheses are <u>not</u> mutually exclusive <u>and</u> exhaustive. In Section 3 we focus on specific problems associated with using the LR on mixed profile DNA samples, namely:

> a) widespread misunderstanding of the potential impact of non-exhaustive hypotheses on the value of the LR, which is especially acute in mixed profile analyses as there will generally be multiple possible 'alternative' hypotheses
> b) the conclusion that the LR result may be essentially meaningless in the absence of prior information (other than the suspect's profile) even if the hypotheses are exhaustive.

In Section 4 we provide an example of the problems using a typical example of an LR analysis based on an anonymized real case.

We refer readers to [27], [28] for details of the formation and general interpretation of epg's and 'routine' statistical calculations involved in DNA profiling of non-complex profiles such as single source or profiles of the type described above (single source, or with a known or identifiable contributor). For special cases, such as those involving relatives - see [29] [30]. A good introduction for mixed profile analysis can be found in [12] [31], [32].

# 2. The likelihood ratio and its limitations as a measure of probative value of evidence

### 2.1. The Likelihood Ratio and Bayes' Theorem

As we receive more information about anything, we update our thinking to accommodate the new information. Simply put, if the new information supports our initial belief then the strength of that belief is increased. The information may not support, but not totally disprove, our belief, in which case we reduce the strength of our belief. Some information may neither support nor refute the belief. Bayes' Theorem provides a method to update our belief in the light of the new information or *evidence*. If our belief was strong before we had the new information then we may take a lot of convincing to significantly change our belief; even though there is some evidence against our belief, it may not be enough to convince us that our belief is false.

Consider a screening test for an infectious virus. The test is considered accurate because it has low false negative and false positive rates. Specifically:

- The false negative rate is just 1% (this means that 1% of people who have the virus will wrongly be diagnosed as negative; equivalently 99% with the virus will be correctly diagnosed as having it)
- The false positive rate is just 2% (this means that 2% of people who do NOT have the virus will wrongly be diagnosed as positive; equivalently 98% of people who do not have the virus will be correctly diagnosed as not having it)

Now suppose you take the test and the result is positive. What does this evidence tell you about your chances of having the virus?

The way forensic scientists are encouraged to answer this question is to consider the ratio of the probability of the evidence if you have the virus and the probability of the evidence if you do not have the virus. This is called the Likelihood Ratio (LR), i.e.

$$LR = \frac{\text{Probability of a positive test if you have the virus}}{\text{Probability of a positive test if you do NOT have the virus}}$$

In this case
$$LR = \frac{0.99}{0.02} = 49.5$$

As the LR is greater than one, the evidence is said to 'provide support' for the hypothesis that you have the virus. In general, the higher the LR the 'stronger' the support. However, no matter how high the LR, it actually tells us nothing directly about the actual probability of the hypothesis. Suppose, for example that recent data shows that the virus currently affects one in two hundred of the population, then it is reasonable to start with the initial assumption that the probability YOU have the virus is 1 in 200 (we call this the 'prior probability'). With this assumption, the correct revised probability (we call this the 'posterior probability') that you have the virus 1/5, i.e. there is a 20% chance. In other words, despite the 'strong evidence' of the positive test result, it is still unlikely you have the virus. Although to many this sounds counterintuitive, Figure 1 demonstrates it is correct graphically.

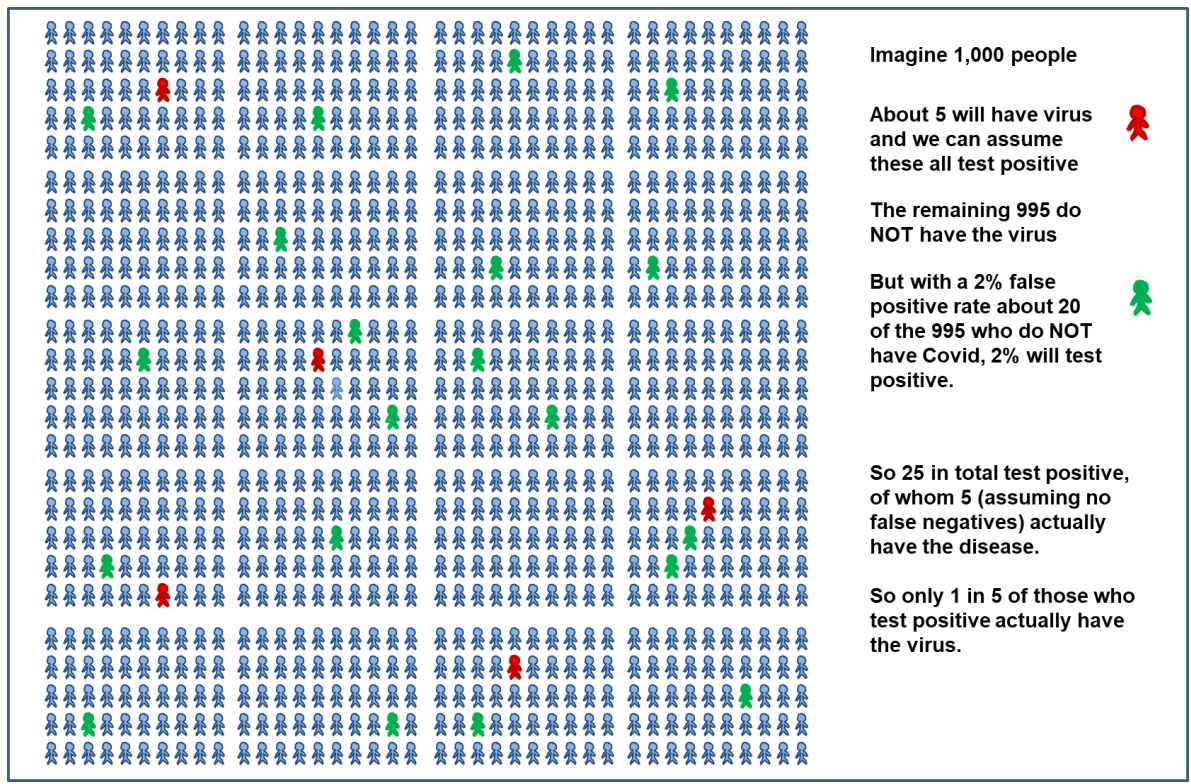

**Figure 1 Graphical explanation of Bayes Theorem**

In fact, it Bayes theorem that provides the formal way of doing this calculation. This is the general formula (See Box 1) for calculating the posterior probability based on the prior probability and the new evidence (which in this case is the positive test result).

> **Box 1: Bayes Theorem**
>
> We have a hypothesis $H$ ("Have the virus") which has a prior probability of 1/200. We write this as $P(H)=0.002$
>
> We get some evidence $E$ ("Positive test result").
>
> We know the probability of $E$ if $H$ is true is 1. We write this as $P(E|H)=1$.
>
> We also know the probability of $E$ if $H$ is false is 0.02. We write this as $P(E|\overline{H}) = 0.02$.
>
> What we want to know is the posterior probability of $H$, that is the probability of $H$ given $E$, written $P(H|E)$.
>
> Bayes theorem provides the formula for this as:
>
> $$P(H|E) = \frac{P(E|H) \times P(H)}{P(E|H) \times P(H) + P(E|\overline{H}) \times P(\overline{H})}$$
>
> So, in our example
>
> $$P(H|E) = \frac{1 \times 0.005}{1 \times 0.005 + 0.02 \times 0.995} = 0.2008$$

It is because of Bayes theorem (and ***only because of Bayes theorem***) that the LR is considered to be a valid measure of probative value of evidence. This is because of what is termed the 'odds form' of Bayes Theorem (see Appendix A1). Instead of framing our belief in a hypothesis as a probability we can equivalently frame it in terms of the odds that our hypothesis is correct. For example, the odds for a hypothesis with probability 1/10 are 11 to 1 against. The odds version of Bayes theorem is a formula for updating the extent to which we change the prior odds based on the strength of the new evidence. The formula tells us simply that the posterior odds are the prior odds multiplied by the LR. So, whatever *prior* odds we start with, an LR greater than 1 will increase the *posterior* odds (because any number multiplied by a number greater than 1 will increase) and an LR less than 1 will decrease it.

## 2.2. The Likelihood Ratio (LR) for mutually exclusive and exhaustive hypotheses

Generally, the LR enables us to compare which of two hypotheses is better supported by some evidence. In the legal context the two hypotheses are the prosecution hypothesis and the defence hypothesis, and the LR is the ratio

$$\frac{\text{Probability of }\textit{evidence}\text{ if prosecution hypothesis is true}}{\text{Probability of }\textit{evidence}\text{ if defence hypothesis is true}}$$

Superficially, the LR seems to be a good measure of 'probative value of the evidence' in the sense that:

- If the LR is greater than 1 then the evidence 'better supports' the prosecution hypothesis (with higher values showing greater support) because the probability of the evidence is greater for the prosecution hypothesis than the defence hypothesis
- If the LR is less than 1 then the evidence 'better supports' the defence hypothesis (with lower values showing greater support)
- If the LR is equal to 1 then the evidence supports both hypotheses equally because the probabilities of the evidence under each hypothesis is the same and hence is 'neutral'.

What makes the LR an especially attractive 'measure' of the strength of evidence for forensic scientists [33], [34] is that it can be calculated without reference to, or consideration of, the prior probability of either hypothesis. Moreover, there is an additional perceived benefit of the LR in that it forces expert witnesses to consider separately the probability of the evidence under both hypotheses.

However, there is much confusion about why and how the LR can really be considered a meaningful 'measure of probative value'. Any reasonable notion for some evidence to have 'probative value' in support of a hypothesis requires our belief in the probability of the hypothesis being true to increase after observing the evidence (i.e. the posterior probability must be greater than the prior probability). Crucially, for the LR to be a meaningful measure of the probative value of the evidence as intended, it is a necessary requirement (as explained formally in Appendix 1.2) that the hypotheses are mutually exclusive and exhaustive – meaning that they are the *only* possibilities. For example, there is little point in having evidence that tells us that the odds of one horse winning a race are better than another horse winning the race unless we know the odds for all of the horses in the race.

This means that the defence hypothesis must simply be the negation of the prosecution hypothesis (e.g. A v *not* A). In this case (the 'odds version' of) Bayes Theorem asserts:

> the posterior odds in favour of the prosecution hypothesis are simply the prior odds multiplied by the LR. (1)

So, if the LR is greater than 1 then the posterior probability of the prosecution hypothesis is greater than its prior probability, i.e. the evidence leads to an increase in the probability that the prosecution hypothesis is true.

Hence, providing that the defence hypothesis is simply '*not* the prosecution hypothesis' (i.e. 'prosecution hypothesis is false'), the LR is a genuine and meaningful measure of the probative value of evidence in the sense originally popularised in [35], since it translates directly to changes in probability of the hypotheses.

While the LR is a valid measure of probative value of evidence for exhaustive hypotheses it cannot tell us the probability of the hypothesis without considering the *prior* probability (i.e. how much did we believe the hypothesis in the first place - how strong does the evidence have to be to change my belief). In other words, we can say how much the probability has increased over the prior, but we cannot conclude anything about what the actual posterior probability of the hypothesis is. When the prior probability of a hypothesis is very low, a high LR in favour of this hypothesis may still result in a low posterior probability that the hypothesis is true. Our virus testing example demonstrated this. Appendix 1.3 provides other simple and extreme examples of this.

Hence, a very high LR does not imply a very high probability of the hypothesis being true. So, while a forensic scientist may conclude, as many do [36], from a very high LR

> "this provides strong support for the prosecution hypotheses"

this does not necessarily mean that the prosecution hypothesis is likely to be true. Unfortunately, a very high LR is commonly reported without discussion of the prior. It is known that lay people may assume this means there is a very high probability that the prosecution hypothesis is true. In other words, they confuse the probability of the *hypothesis given the evidence* (the posterior probability) and the probability of the *evidence given the hypothesis*. Assuming these are the same is a frequent error called the transposed conditional fallacy, or the 'prosecutors fallacy' [37], [38]. Hence, without very careful explanation, the use of such a LR in court has the potential to bias the jury [39].

## 2.3. The Likelihood Ratio (LR) for non-exhaustive hypotheses

So, unfortunately, as shown in Appendix 1, it is *only* when the two hypotheses are mutually exclusive and exhaustive that the LR has genuine meaning as a measure of probative value of evidence. When the hypotheses are not exhaustive (i.e. when the defence hypothesis is not simply the negation of the prosecution hypothesis) the LR is not a measure of 'probative value' of the prosecution hypothesis (as many wrongly assume). In this case Bayes Theorem is:

> the posterior odds of the prosecution hypothesis against the defence hypothesis are equal to the LR times the prior odds of the prosecution hypothesis against the defence hypothesis  (2) [also see Appendix A1]

Note that, in contrast to theorem (1) where we had exhaustive hypotheses, Bayes does not allow us to conclude anything about the posterior probability of the prosecution hypothesis in this case. All we can conclude from a LR>1 (no matter how big it might be) is that the relative odds of the evidence of the prosecution to defence hypothesis have shifted in favour of the prosecution hypothesis. But the actual probability of the prosecution hypothesis may have *decreased* rather than increased, so in no sense does it 'support the prosecution hypothesis'. Indeed, Appendix 2 provides examples that prove the following are possible when we have non-exhaustive hypotheses:

- LR >1 but probability of the prosecution hypothesis decreases. So LR>1 but the evidence does NOT support the prosecution hypothesis.
- LR=1 but the posterior probability of the prosecution hypothesis increases. So LR=1 is not neutral, as the evidence supports the prosecution hypothesis.

So, suppose the evidence is "DNA found at the crime scene whose profile matches the defendant's DNA". A typical prosecution hypothesis in such a case would be:

> "the DNA is from the defendant".

In order to use the LR meaningfully the defence hypothesis must be

> "the DNA is NOT from the defendant".

However, in practice forensic scientists typically choose an alternative defence hypothesis that is not the negation, namely:

> "the DNA is from a person unrelated to the defendant".

They do this because it is statistically simpler to provide an estimate for this likelihood. This standard assumption means that any LRs used for DNA match evidence are based on non-exhaustive hypotheses and are therefore not a measure of probative value of the prosecution hypothesis.

Because relatives who could have left the DNA are ignored, it is possible that the evidence may be a better match for some close relative than it is for the suspect. This is especially relevant for LTDNA samples where there may be uncertainty about what the alleles are on several loci. In such a case, although the LR will be very high it may not actually 'support' the prosecution hypothesis. Unfortunately, using the verbal scale [40], [41], it is very common [10] for forensic scientists to make the following assertion in such cases about the high LR:

> 'this provides strong [or extremely strong] support for the prosecution hypothesis'.

Such an assertion is not correct and is unfairly prejudicial against the defendant. What they should conclude is:

> 'this appears to provide support for the prosecution hypothesis over this *particular* defence hypothesis but may not provide support that the prosecution hypothesis is more likely to be true'.

By allowing non-exhaustive hypotheses, the defence could also cherry-pick alternative hypotheses in *their* favour. For example, suppose the defendant is known to have a twin brother and the defence lawyers insist that their hypothesis is that "the DNA is from the twin brother" (thereby ignoring the possibility the DNA is from a person unrelated to the defendant). In this case, the LR would be equal to 1 and the defence lawyer might therefore claim that the evidence has no probative value. However, this is unfairly prejudicial against the prosecution case; the evidence does have probative value in support of the prosecution hypothesis because the posterior probability that the DNA is from the defendant will certainly increase [42].

The problem of non-exhaustive hypotheses is even greater when the evidence is a mixed DNA profile, as in such cases there are normally two ways in which the hypotheses used are non-exhaustive:

1. Because there is uncertainty about the number of contributors to the mixture, the hypothesis 'suspect is a contributor to the mixture' will inevitably be conditioned on an assumption such as 'there are $n$ contributors' where $n$ is a number greater than 1
2. The typical alternative hypothesis is: 'There are $n$ contributors all unrelated to the suspect' rather than 'The suspect is not one of the $n$ contributors'

In such situations, the LR may be misleading because of failure to consider related contributors, as well as failure to consider a different possible number of contributors. Indeed, changing the assumption about number of contributors usually drastically changes the LR (for example, a 3-person assumption might produce a LR close to 0, i.e. strongly favouring exclusion of the suspect, whereas a 4-person assumption might produce a very high LR favouring inclusion). So, the LR for mixed profile DNA evidence is already potentially misleading because of built in non-exhaustivity. But, as we will show in Section 3, there are even more fundamental (but not widely recognised) reasons to be wary of the LR for mixed profile DNA evidence.

### 2.4. Other general concerns about the Likelihood Ratio

The above points (i.e the importance of exhaustivity for the LR to be a measure of probative value and the need for a prior in order to use the LR to make conclusions about the posterior probability) have been long recognized by multiple statisticians and forensic scientists [43]–[51]. Even in a highly relevant legal paper [52] that reviews the use of Bayes and the LR in legal cases, the author asserts (page 3) that "propositions are mutually exclusive and exhaustive" (where 'propositions' refer to the prosecution and defence hypotheses).

Despite this, enormous confusion continues in practice. An indication of the extent of the confusion can be found in one of the many responses by the statistics and forensic scientist community to the RvT judgement [53] (a ruling which rejected the use of Bayes and the LR for any forensic match evidence other than DNA, after a bungled attempt to use it for footwear match evidence). Specifically, in an otherwise excellent position statement [54] (signed by multiple experts) is the extraordinary point 9 that asserts:

> "It is regrettable that the judgment confuses the Bayesian approach with the use of Bayes' Theorem. The Bayesian approach does not necessarily involve the use of Bayes' Theorem."

By the "Bayesian approach" the authors are specifically referring to the use of the LR, thereby implying that the use of the LR is appropriate, while the use of Bayes' Theorem may not be. As explained in Sections 2.1 and 2.2 this assertion misses the core point that the LR - as a measure of probative value of evidence - is only meaningful *because* of Bayes Theorem. This was made clear by Equation (1) for exhaustive hypotheses and equation (2) for non-exhaustive hypotheses.

So, the idea that using the LR is 'not using Bayes theorem' is a misconception. If a forensic scientist was questioned in court about why a high LR was probative in support of the prosecution hypothesis or probative in support of the prosecution hypothesis over a non-exhaustive defence hypothesis, they would be forced to resort to Bayes for an explanation. There is no other explanation.

Further concerns about the use and limitations of the LR as a measure of 'weight of evidence' were detailed in [51] and in [55]. Specifically:

- Even when hypotheses are mutually exclusive and exhaustive, there remains the potential during a case to confuse source-level hypotheses (such as DNA on a weapon found belonging to or not belonging to the defendant) and activity-level hypotheses (such as defendant was at the crime scene or not) or even offence-level hypotheses (such as defendant being guilty or not guilty). Sometimes one may mutate into another through slight changes in the precision with which they are expressed. A LR for the source-level hypotheses will not in general be the same as for the activity-level or offence-level hypotheses. Indeed, it is possible that an LR that strongly favours one side for the source-level hypotheses can actually strongly favour the other side for the activity-level hypotheses even though both pairs of hypotheses seem very similar. Similarly, an LR that is neutral under the source-level hypotheses may be significantly non-neutral under the associated activity-level hypotheses.

- In many situations the LR represents an over-simplification of the underlying hypotheses and evidence. For example, a defence hypothesis "defendant is not the source of the DNA found" is, in reality, made up of multiple hypotheses that are difficult to articulate and quantify (technically every person in the world). As we already noted earlier, the standard pragmatic solution (which has been widely criticised [43]) is to assume that the hypothesis represents a 'random person unrelated to the defendant'. But not only does this raise concerns about the homogeneity of the population used for the random match probabilities, it also requires separate assumptions about the extent to which relatives can be ruled out as suspects. Also, it is not just the hypotheses that may need to be 'decomposed'. In practice, even an apparently 'single' piece of evidence E comprises multiple separate pieces of evidence, and it is only when the likelihoods of these separate pieces of evidence are considered that correct conclusions about probative value of the evidence can be made. Often, to avoid this, grossly simplified assumptions have to be made in order for the LR to be computed without specialist software. Moreover, even where software (such as probabilistic genotyping software) is used it still makes grossly simplified assumptions about independence between components of evidence (to do things properly Bayesian network software must be used but rarely is).

Many authors have raised specific concerns about the use of the LR for DNA analysis [56]. For example, [57] assert:

".. the numerical value of the likelihood ratio depends on the particular choice of the hypotheses used by the expert. Since the choice of the hypotheses is sometimes arbitrary, this implies that reporting only the likelihood ratio (in conjunction with the hypotheses of course) as a measure for the strength of the evidence can lead to misinterpretations".

The authors of [57] conclude that the LR should not be presented without consideration of the effects of the prior odds. In response to this article Balding identifies further limitations of the LR for DNA analysis [43]. In particular, he questions the use of alternative hypotheses based on a 'random man' (i.e. "other unknown" person) arguing that it is necessary to consider alternative hypotheses for "every other individual alternative culprit". In another response [50] confirm Balding's concerns asserting that, even for a single DNA profile, there should really be "about 6,000,000,000 hypotheses, one for each person on earth".

# 3. The particular problem when using LRs for DNA mixture profiles in contrast to single profiles

We already identified that the use of the LR for mixed profile DNA evidence was potentially even more misleading than the LR for single profile DNA evidence because, in addition to the assumptions about unrelated contributors, there are also assumptions about the number of contributors. In other words, significant 'non-exhaustive' hypotheses issues are present. But things are much worse still. To explain why, we first go back to some basics for those unfamiliar with DNA profiling and then expose the special problems of using the LR for DNA mixtures. Finally, we show that probabilistic genotyping, i.e. using continuous rather than binary allele identification, does not resolve fundamental problems of the meaningfulness of the LR for mixed profiles.

### 3.1. DNA profiling basics

At each locus of a DNA profile there are short sequences of base pairs repeated multiple times – these are called ***Short Tandem Repeats*** (STRs). The number of times that the sequences are repeated varies between individuals. The length of each repeated sequence can be measured and expressed as the number of repeats in the sequence. This is called an *allele*. At each locus there are two alleles - one inherited from the father and one from the mother. So, a person's DNA profile is just a sequence of number pairs like:

| Locus | D3 | D8 | D18 | D19 | D21 | vW1 | TH01 | FGA | … |
|---|---|---|---|---|---|---|---|---|---|
| Allele | 14,15 | 13,13 | 12,15 | 14,16 | 29,29 | 16,16 | 6,7 | 24,25 | … |

The pair of alleles is called the *genotype* of the person at that locus. The pair may be the same, such as in the loci D8 and vW1 above, in which case we say the genotype is *homozygous*. Otherwise we say the genotype is *heterozygous*. A person's DNA profile is analysed typically from a sample of their saliva, and a machine provides a printout in which the alleles are identified by 'peaks' such as shown here:

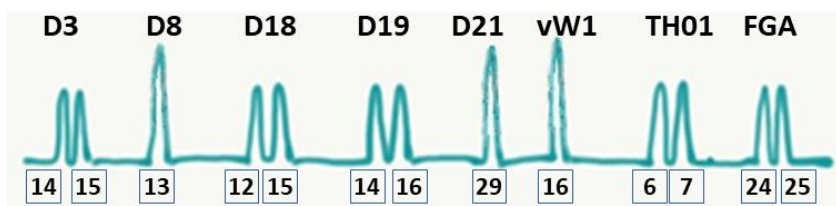

Generally, a DNA sample found at a crime scene will be determined to be either a single profile (meaning it comes from a single person) or mixed (meaning it comes from at least two people). If, for example, at every loci there are no more than two peaks (i.e. two alleles identified) then it is highly likely that it is a single profile (although, because of the possibility of multiple homozygous genotypes, it is theoretically possible that the profile is mixed). In the case where it is clear that it is a single profile of good quality, then the problem of determining a 'match' against a suspect's profile is relatively straightforward - if we ignore the issues raised above in assuming that the defence hypothesis is that the DNA came from an 'person unrelated to the suspect'. We simply have to check that, for each locus where there is an identified genotype, the suspect's genotype is the same at that locus. In such situations the numerator of the LR will typically be 1 and the denominator will be the probability that a person unrelated to the suspect has the same DNA profile (this is often referred to as the 'random match probability'). Assuming no errors are made in the analysis, it is accepted that the suspect's DNA matches the profile of the source DNA; although this does not mean the suspect is the source of the DNA, a high LR genuinely means it is highly unlikely that an unrelated person would have the same matching profile. Indeed, in this formulation, the LR is simply the inverse of the random match probability (RMP) and so the RMP can be easily inferred from the LR.

Unfortunately, things become very different in the case of a mixed DNA profile, i.e. one where there are at least 3 different alleles at more than one locus.

### 3.2. Why things are very different for mixed profiles

For a start, in a mixed profile there is no certainty about the number of different contributors to the mixture (although methods can be used to determine the probabilities of any given number). As the number of contributors increases the probability of underestimating the true number also increases [11], [58], [59]. Moreover, assuming there are at least two unknown contributors (so we are not considering the case where there are two contributors one of whom is known to be the victim) there can never be certainty about whether a suspect's DNA profile is actually in the mixture at all [60]. To see why, consider the following mixture profile:

| Locus | D3S1358 | vWA | D16S539 | …. |
|---|---|---|---|---|
| **Alleles found** | 14, 15, 16 | 16, 17, 18 | 9,10,11, 12 | … |

Now suppose the suspect's DNA profile is

| Locus | D3S1358 | vWA | D16S539 | …. |
|---|---|---|---|---|
| **Suspect alleles** | 15, 16 | 17, 17 | 9,11 | … |

Then the suspect's alleles are certainly included in the set of alleles at each locus. However, even if we could assume there are exactly two contributors, then there are still multiple pairs of possible profiles **that would exclude the suspect from being a contributor**. For example, all the following possibilities involve a pair of contributors where *neither* matches the suspect on *any* locus:

| Locus | D3S1358 | vWA | D16S539 | …. |
|---|---|---|---|---|
| **Contributor 1** | 14, 16 | 16, 17 | 9,10 | … |
| **Contributor 2** | 14, 15 | 17, 18 | 11,12 | |

| **Contributor 1** | 14, 16 | 16, 17 | 9,12 | … |
|---|---|---|---|---|
| **Contributor 2** | 15, 15 | 18, 18 | 10, 11 | |

| **Contributor 1** | 14, 15 | 16, 16 | 9,12 | … |
|---|---|---|---|---|
| **Contributor 2** | 16, 16 | 17, 18 | 10, 11 | |

There are many more such combinations. Nevertheless, the LR for the hypothesis 'suspect is a contributor' against the hypothesis 'suspect is not a contributor' will be very high, and an analyst would typically conclude

> "this provides strong support for inclusion"

In contrast to the single profile case, this is exceptionally misleading. Even if it certain that a person's alleles are included on every locus of a mixed DNA profile then, while there will be a very high LR, the probability that the person is a contributor (in the absence of any other evidence) is actually very low. This is because, as we have explained above, there are many combinations of profiles that do not include the suspect's. Hence, in the absence of any other evidence, the probability for the defendant's inclusion in a mixed profile will be exceptionally low. While this problem has been recognized informally [60] Appendices 3-5 provide a novel formalisation of how serious the problem is with respect to the use of the LR.

Specifically, Appendix 3 shows formally that the following is possible (in a hypothetically simplified scenario) for a mixture with two unknown contributors:

> **Without strong prior assumptions about the probability that a suspect is a contributor** to a 2-person mixture, the LR for the (standard) hypothesis:
>
> > "suspect is a contributor"
>
> (even assuming the alternative hypothesis is its negation) **is essentially meaningless**. In contrast, the LR for the hypothesis
>
> > "A DNA profile matching the suspect's profile is contained in the mixture"
>
> is meaningful without any prior information. However, in this case, despite a high LR, the posterior probability will be low.

Appendix 4 further exposes the extreme limitations of what can be deduced from a very high LR when there are at least two unknown contributors and the hypothesis is "suspect is a contributor" but there is no other evidence against the suspect. Using a simple analogy with numbered balls we show that, in theory, if the suspect's alleles are included on every locus of a mixed profile then this will certainly lead to an astronomically high LR, but (in stark contrast to the single profile match scenario) the posterior probability that the person is a contributor is actually very low.

Appendix 5 explains how it is possible for a mixture profile to have multiple different prosecution hypotheses for the same evidence, all with astronomically high LRs, even though at most one of the hypotheses is true and the others are false. For example, if we do not know the number of contributors to the mixed profile, then among the many possible prosecution hypotheses we get:

1. By assuming 3 contributors the prosecution can claim: "The evidence is $1.1*10^{12}$ (more than a trillion) times more likely if suspect is a contributor than if he is not."

2. By assuming 4 contributors the prosecution can claim: "The evidence is $1.73*10^{10}$ (more than 10 billion) times more likely if suspect is a contributor than if he is not."

But at least one of the hypotheses: "Suspect + 2 unknowns" or "Suspect + 3 unknowns" must be false. Without loss of generality suppose the first is false. Then it follows that, despite the fact that the evidence is more than a trillion times more likely if a Fred is a contributor than if he is not, it is CERTAIN than Fred is not actually a contributor to a 3-person mixture.

So we have multiple different prosecution hypotheses (at most one of which can be true) all yielding incredibly high LRs.

It is also worth pointing out that, while the results in Appendices 3- 5 use only theoretical examples, it turns out that in practice, LRs of greater than one for ***known non-contributors*** to mixture profiles are observed [14], [61], [62].

### 3.3. Does probabilistic genotyping address any of these concerns?

As explained in Section 1, many now believe that for mixture DNA profiles – especially low template samples – continuous, rather than binary, inclusion criteria, should be used in any LR calculations. This potentially makes it possible to take account of the heights of each allele peak which in turn makes it possible to distinguish between 'major' and 'minor' contributors and also to potentially accommodate the probability of drop-out, stutter and other errors in low-template samples that can lead to alleles being wrongly determined as absent/present. The numerous probabilistic genotyping software systems (mentioned in Section 1) that use continuous inclusion criteria claim to be able to properly deal with drop-out, stutter and other errors. These systems also automatically compute LRs for mixed profiles.

In practice the probabilistic approach can address some of the concerns described in Section 3.2 because, in practice, different contributors may contribute different amounts of DNA, which is reflected in different peak heights for their alleles [63]. In such cases some of the problems highlighted in Section 3.2 are less relevant because much of the uncertainty about whether genotypes matching the suspect's are actually in the mixture at all is reduced and some genotypes can be excluded on the basis of the peak heights [3].

However, in many real crime instances it is common to find a mixture in which a known victim contributes the major component (e.g. 80%) while 2 or more unknown contributors all have a similar sized minor component (e.g. 10% each for 2 contributors). Although the peak heights can still assign probabilities to whether or not borderline alleles are present or not, they do not help differentiate between the contributors. Hence, the problems associated with the LR for whether a suspect is one of the unknown contributors in such cases is exactly the scenario described in the examples in Appendices 3-5. In other words, in such situations the continuous rather than binary approach still results in LRs of questionable value.

So, in the worst case where two unknown contributors of 'equal size' are in a mixture there will be many billions of different possible DNA profiles that could be in the mixture (i.e. are not excluded). For **every** such profile X, the LR for the hypothesis "X is in the mixture" against "X is not in the mixture" will be an astronomically high number. Yet, despite this high LR, the profile X is no more likely to be in the mixture than any of the other billions not excluded. The problem with the standard mixed profile analysis process is that it is typically subject to confirmation bias [64]. Instead of being asked to determine which two (of the billions of) profiles are most likely to make up the mixture, the analysts may typically be asked to determine the LR for a given profile X being in the mixture. And, inevitably, the LR is astronomically high. When presented with this LR lay people are unaware that the same astronomically high LR would have been produced for any of the other billions of profiles not excluded from the mixture.

The examples in Appendix 3-5 assume exhaustive hypotheses and so in practice the problems are even worse than stated there, because it is clear that it has become almost standard to consider hypotheses that are not exhaustive in the LR calculations for mixture profiles.

The normal justification for using non exhaustive hypotheses in reporting LRs for mixture profiles is an underlying assumption that the hypotheses chosen represent the only 'realistic' alternatives. But, as the hypotheses generally exclude the possibility of related contributors, this means that any LR in favour of inclusion of the defendant will be systematically overestimated [65]. Even assuming no related contributors there are generally multiple alternative hypotheses based on the different possible number of contributors. An often recommended method for dealing with such multiple alternatives is to produce a likelihood calculation for each (i.e. the probability of the evidence if that particular hypothesis is true) instead of a LR for just two hypotheses. In such an analysis the hypothesis with the largest likelihood is the one best supported by the evidence. However, it seems this is rarely done and, in any case, still tells us nothing about the probability of that hypothesis being true.

### 3.4. How reliable are the probabilistic genotyping software systems

The complexity, limitations and different results that arise from the probabilistic genotyping software products are widely acknowledged and discussed in papers such as [10] and [65] with the latter asserting:

> "Although several methods have been developed, they each suffer from limitations of various kinds and none of the methods have been universally acknowledged as gold standards."

The reference [65] also includes multiple responses from experts in the area that highlight disagreements.

A major concern with the LR values reported by specific software products is that they are extremely sensitive to apparently trivial and irrelevant changes to the DNA amplification and electrophoresis process; specifically, small changes in this process can result in LRs for the same sample that differ by several orders of magnitude [26]. Also, different software products can produce orders of magnitude different results for the same sample [24], [25], (although some claim the scale of the problem is exaggerated [32], [66], [67]). Hence, the LR values reported may not be reliable or accurate. Suggestions that the reproducibility of results could be tested by sending the same known samples to several laboratories and comparing the results after undergoing the entire laboratory profiling process [68] have been rejected by some software developers [69].

To take proper account of the complexity of mixed profile probabilistic calculation, the authors of [65] and also [70] use Bayesian network models and software to calculate LRs (something we believe to be necessary) but none of the better-known software products incorporate such methods.

It is also important to note some concerns about the way the different software systems handle certain problematic alleles in LTDNA samples. Unlike binary methods, continuous methods assign probabilities to an allele being present/absent at a locus even though the peak height may fall below/above the pre-defined threshold for reporting. In such situations the probabilities are purportedly taking account of dropin, dropout and stutter. Hence, it is quite common for a person to be considered a possible contributor to a LTDNA sample even though their alleles are formally reported as absent from some loci in the analysis (we will see examples of this in Section 4). As described in [29] such anomalies have been referred to as 'voids' [71]. Indeed, it is not uncommon for astronomically high LRs to be reported in support of inclusion even though the suspect's alleles are reported as missing at several loci. A major motivation in [29] was to show that treating voids as neutral was unfavourable to the defendant; yet although their approach does lead to lower LR's in favour of inclusion in such cases, it may still be unfavourable to the defendant. In particular, [29] assume, based on the paper [5] that:

- drop-out is independent across loci,

- drop-ins at different loci are mutually independent, and independent of any drop-outs.

These independence assumptions may be unrealistic but are made so that the LR calculations can be done relatively easily. A model with possibly more realistic assumptions is one in which there is a common unknown 'error rate' variable - which can have a fairly strong 'low' prior - but which will 'increase' the more voids we see in a particular profile analysis.

## 4. Example LR result presentation

Table 1 presents a snapshot from the output a probabilistic genotyping system, along with the summary results and conclusion reported. This is based on a real example (with some details changed to ensure anonymity).

**Table 1 Sample output from mixed profile DNA analysis**

| Locus | Alleles in sample | Suspect alleles |
|---|---|---|
| D3S1358 | 14, 15, 16 | 15, 15 |
| vWA | 15, 16, 17, 19 | 17, 19 |
| D16S539 | 10, 11, 12 | 12, 12 |
| D8S1179 | 11, 12, 13, 14 | 11, 12 |
| D18S51 | 12, 14, 17, 25 | 13, 25 |
| … | … | … |
| D5S818 | 7, 10, 11, 12, 13 | 10, 11 |
| D13S317 | 8, 9, 10, 12, 13 | 9, 10 |
| D7S820 | 8, 8.3, 9, 9.3, 10, 12 | 10, 11 |
| … | … | … |
| D12S391 | 15, 16, 18, 19 | 18, 18 |
| D2S1338 | 16, 17, 19, 20, 24, 25 | 23, 25 |
| **RESULTS OF ANALYSIS** | | |
| **Assumed number of contributors**: 4 | | |
| **Likelihood Ratio**: It is $4 \times 10^6$ times more likely to obtain the DNA results if suspect is a contributor than if he is not a contributor | | |
| **Summary conclusion**: Strong support for inclusion of suspect | | |

The highlighted rows indicate there is a void at this locus, i.e. these are instances where the suspect's genotype is not included in the mixture profile. However, the probabilistic genotyping analysis takes full account of the peak heights of each allele in the mixture and the possibility of dropout and dropin so that, for example, the probability that allele 13 is 'missing' in locus D18S51 due to dropout is incorporated. Nevertheless, the extremely high LR 'supporting inclusion' of the suspect is highly counter-intuitive given the fact there were 4 loci at which the suspect's genotypes were not detected. On top of this problem, based on the discussions in Sections 2 and 3, the analysis and conclusion are misleading even if we accept the multiple assumptions made for the statistical analysis (and assume the DNA was processed without errors):

1. *We cannot make any conclusions about the suspect being a contributor from this LR without assuming some prior probability for suspect being a contributor*. In particular, even if it certain that a person's alleles are included on every locus of a mixed DNA profile then, while there will be a very high LR, the probability that the person is a contributor (in the absence of any other evidence) is actually very low.

2. *The prosecution and defence hypotheses are not exhaustive*. There could, for example, have been a different number of contributors, and it is also possible that there were unknown contributors related to the suspect. All we can actually conclude is that: 'it is $4 \times 10^6$ times more likely to obtain the DNA results if the suspect and three unknown (and unrelated) people are contributors than if there are four unknown (and unrelated) contributors'. This tells us nothing about the probability that the suspect is a contributor, and it is possible that some other hypothesis is better explained by the evidence than "suspect and three contributors". Indeed, because the suspect's alleles were not detected at several loci it is possible that there are people related to the suspect who are more likely to have been contributors.

In the light of the above, the DNA expert *should* have reported the evidence as follows:

> "If we could be certain that there were exactly 4 contributors to this mixture, then it is $4 \times 10^6$ times more likely to obtain the DNA results if the suspect is a contributor than if four people unrelated to the suspect were the contributors. However, we cannot be certain that there were four contributors; if there were a different number then it is possible that the suspect could be definitively excluded. Also, because the suspect's alleles were not detected at several loci, it is possible that there are people related to the suspect who are more likely to have been contributors. Even if we are sure of the number of contributors and can exclude the possibility of people related to the suspect being contributors then, in the absence of any other evidence linking the suspect to the DNA sample, it is still very unlikely that he is a contributor."

## Summary

By using probabilistic genotyping software, a DNA expert will typically report the results of a mixed DNA sample by citing a LR ratio (X) as follows:

> "It is X times more likely to obtain these DNA results if the suspect is a contributor than if he is not a contributor. Hence these results provide strong support for inclusion of the suspect"

When X is of the order of billions or even trillions, lay people are known to interpret this result as conclusive evidence that the suspect is a contributor to the sample. We have shown, however, that reporting the results in this way may be highly misleading for the following reasons:

1. Even if we assume that the hypotheses are mutually exclusive and exhaustive it is possible in theory (as demonstrated in Appendix 3 and 4) to achieve astronomically high LRs even when it is almost certain that the suspect is NOT a contributor. This applies in scenarios where there are at least two unknown contributors of similar 'size'. Indeed, in the absence of any prior information, the posterior probability of inclusion in certain scenarios will always be very low. Moreover, we also showed that without strong prior assumptions about the probability that a suspect is a contributor, the LR for the (standard) hypothesis: "suspect is a contributor" in such scenarios is essentially meaningless and should be replaced with the hypothesis

    > "A DNA profile matching the suspect's profile is contained in the mixture"

    This latter hypothesis is meaningful without any prior information. But again, in this case, it is possible to have an astronomically high LR yet still have a very low posterior probability.

2. The prosecution and defence hypothesis are rarely (if ever) exhaustive. In contrast to what is implied, the defence hypothesis used in the LR calculation is NOT 'suspect is not a contributor' but is rather subset of the scenarios under which the suspect is not a contributor, namely those in which: 1) there is an assumed fixed number of contributors even though this number cannot be known with certainty;

and 2) all the contributors are assumed to be unrelated to the suspect. Hence possibly better alternative 'explanations' for the evidence are entirely excluded.
3. Because the hypotheses are not mutually exclusive and exhaustive, the LR tells us nothing about how the evidence changes the probability of the prosecution hypothesis being true. In theory a very high LR can even result in a reduced probability of the prosecution hypothesis being true if there is an alternative hypothesis not included as part of the defence hypothesis.
4. For LTDNA mixed samples it is common for some of the suspect's alleles to be missing at a number of loci. Because probabilistic genotyping methods account for the possibility of such 'voids' being due to dropout, very high LRs in support of inclusion of suspect are still reported.
5. Even if we assume that the hypotheses are mutually exclusive and exhaustive, we can conclude nothing about the probability of the prosecution hypothesis without assuming the prior probability of the prosecution hypothesis. However, in this case a high LR *does* tell us that the probability of the prosecution hypothesis has increased (and specifically, by Bayes, the odds in favour of the prosecution hypothesis have increased by a factor equal to the LR). But, if the prior probability is very low (as it will be for a hypothesis about the defendant's inclusion in a mixed profile in the absence of other evidence) a very high LR may still result in a low posterior probability of the prosecution hypothesis.

In cases where there is only one unknown contributor, or where each unknown contributor contributes very different amounts of DNA (reflected in different peak heights for their alleles) some of the above problems are alleviated when using the probabilistic (i.e continuous allele identification) approach. In such cases much of the uncertainty about whether genotypes matching the suspect's are actually in the mixture at all is reduced. However, analysts have been interpreting such cases for decades using what are termed 'restricted' versions of other statistical methods such as the restricted Combined Probability of Inclusion whereby some genotypes are excluded on the basis of consideration of peak heights [14].

There are additional concerns about the reliability of the results of probabilistic genotyping software:

- LR values reported are extremely sensitive to multiple different ways in which the DNA analysis is performed. Different methods (and indeed different software) produce orders of magnitude different results for the same sample.
- The independence assumptions used for handling drop-in and drop-out errors are unrealistic, resulting in LRs that are too high in cases where there are several 'voids' (namely 4 loci for which the defendant's alleles are not shown as present).

In the light of the concerns raised in this paper, we recommend extreme caution when using LR-based LTDNA mixed profile evidence from probabilistic genotyping software. Of special concern would be the following evidence scenario:

A high LR is reported for the suspect being included in a LTDNA mixture where there are at least two unknown minor contributors of similar 'size', one of whom is claimed to be the suspect.

Without other strong evidence linking the suspect to the DNA sample, this evidence is essentially useless because there are typically millions of alternative genotype combinations (none of which are the same as the suspect's) that are equally likely to produce the same mixture (and which would produce equally high LRs). If people have been convicted primarily on the basis of such evidence, then it is likely that there has been a miscarriage of justice.

# Appendix 1: 'Odds' form of Bayes Theorem, the importance of exhaustive hypotheses for the Likelihood Ratio (LR) and the.

**A1.1 Odds form Bayes Theorem**

Figure 6 explains the 'odds form' of Bayes Theorem which 'proves' that the posterior odds of the hypothesis are simply the prior odds times the LR.

### 'Odds' Form of Bayes Theorem

Bayes theorem tells us that:
$$P(H \mid E) = \frac{P(E \mid H) \times P(H)}{P(E)}$$

But equally it tells us that:
$$P(\text{not } H \mid E) = \frac{P(E \mid \text{not } H) \times P(\text{not } H)}{P(E)}$$

If we simply divide the first equation by the second we get the following

$$\underbrace{\frac{P(H \mid E)}{P(\text{not } H \mid E)}}_{\textit{Posterior odds}} = \frac{P(E \mid H) \times P(H)}{P(\text{not } H \mid E) \times P(\text{not } H)} = \underbrace{\frac{P(E \mid H)}{P(E \mid \text{not } H)}}_{= \textit{Likelihood ratio}} \times \underbrace{\frac{P(H)}{P(\text{not } H)}}_{\textit{Prior odds}}$$

**Figure A1 Odds form of Bayes**

We can also use Bayes theorem, as we do in Figure 6, to show that, for any pair of hypotheses H and H' the posterior odds of H to H' are equal to the LR times the prior odds. However, it turns out that it is only when H' is the negation of H that the LR is a meaningful measure of of probative value.

**A1.2 The LR as a measure of probative value of evidence**

Figure A2 explains formally why the notion of the LR as a measure of 'support for a hypothesis' is only valid when the LR involves mutually exclusive and exhaustive hypotheses.

> Evidence 'supports' H if the posterior probability of H is greater than the prior probability of H, i.e. P(H | E) > P(H)
>
> It turns out that this is true precisely when the Likelihood ratio > 1
>
> We prove this from the Odds form of Bayes: $\dfrac{P(H\mid E)}{P(\text{not } H \mid E)} = LR \times \dfrac{P(H)}{P(\text{not } H)}$
>
> If LR > 1, then: $\dfrac{P(H\mid E)}{P(\text{not } H \mid E)} > \dfrac{P(H)}{P(\text{not } H)}$
>
> But: $P(\text{not } H) = 1 - P(H)$ and $P(\text{not } H \mid E) = 1 - P(H\mid E)$
> so substituting these into the above equation we get:
>
> $P(H\mid E) \times (1 - P(H)) > P(H) \times (1 - P(H\mid E))$ which means
> $P(H\mid E) - P(H) \times (H\mid E) > P(H) - P(H) \times P(H\mid E)$ which means
> $P(H\mid E) > P(H)$
>
> **Note that this works only because of Bayes and because the LR involves the mutually exclusive and exhaustive hypotheses *H* and *not H***

**Figure A2 Proof of why the notion of the LR as a measure of 'support for a hypothesis' is only valid when the LR involves mutually exclusive and exhaustive hypotheses.**

### A1.3 When a very high LR still results in a very low probability the hypothesis is true

In the virus example of Section 2 we saw a LR of 59.5 – strongly supporting the hypothesis but the (posterior) probability the hypothesis was true was still very low (about 0.02, i.e. 2%). In fact it is not difficult to come up with examples of much higher LRs that still

Suppose we know that every person is given a number of balls labelled ball1, ball2, ball3 and that each ball contains a random number from 0 to 9. Suppose Fred's balls respectively have the numbers

   1, 7, 0

We will call this sequence "Fred's number profile".

Now suppose that one unknown person left their balls in a bar and that upon inspection it was found that the sequence was 1, 7, 0 (i.e. ball1 was a 1, ball2 was a 7, and ball3 was a 0). Can we conclude that Fred was the person who left his balls there? The LR is easy to calculate:

Let H be the hypothesis that these balls belong to Fred and let E be the evidence that the balls match Fred's. Then:

- The probability of E given H is 1.
- The probability of E given not H is 1/1000 (because there are 1000 different possible number profiles ranging from 0,0,0 to 9,9,9, so the probability that a person other than Fred has exactly the same number profile as Fred is 1/1000

Hence, the LR is 1/(1/1000) which is equal to 1000

However, in the absence of any other evidence, Fred is still no more likely to have left the balls than any other person in the world with the same number profile (the prior probability could be considered as one divided by the number of people in the world). Suppose that, in principle, 10 million people could have been in that bar. Then (either by using Bayes or by noting that approximately 10,000 out of 10 million people would have the same matching profile) it follows that the (posterior) probability that Fred was the person who left the ball is about 1 in 10,000.

We can extend this example to more balls, and this is where some of the type of confusion relating to LRs for DNA evidence arises. If each person had 12 balls then the probability that a person other than Fred has the same number sequence is $10^{-12}$ (that is one divided by a thousand billion); so if that particular number sequence was discovered left in the bar, then the LR for the hypothesis that Fred left the balls is $10^{12}$, i.e. a thousand billion. As there are only about 8 billion people in the world, even if none of them can be ruled out in advance, the posterior probability that it was Fred who left the balls is about 99.2%. For single profile DNA matches it is common to get LRs of this magnitude, and that is why we can normally conclude it is highly likely that the suspect with matching DNA is the person who left it. **However, as we will explain in Appendix 3 and 4, things are completely different for mixed profile DNA 'matches' because these are not 'matches' in the same sense at all; in such cases very high LRs do not even make it likely that the suspect's profile is even included in the mixture**.

# Appendix 2: Problems with the LR when hypotheses are not exhaustive.

### A2.1: When a LR of greater than one in favour of the prosecution hypothesis can actually mean the prosecution hypothesis is LESS likely to be true.

When the prosecution and defence hypotheses are not exhaustive the LR can *only* help us to distinguish between which of the two hypotheses is better supported by the evidence. Unlike the case for exhaustive hypotheses **a LR greater than 1 does not necessarily mean that the evidence 'supports' the prosecution hypothesis**. In fact, the LR can be very large - i.e. the evidence strongly supports the prosecution hypothesis over the defence hypothesis - *even though the posterior probability of the prosecution hypothesis goes down*. This rather worrying point is not understood by all forensic scientists (or indeed by all statisticians). Consider the following example (it's a made-up lottery example, but has the advantage that the numbers are indisputable):

> A lottery has 10 tickets numbered 1 to 10
>
> Joe buys 3 tickets and gets numbers 3, 4 and 5
> Jane buys 2 tickets and gets numbers 1 and 6
>
> The winning ticket is drawn but is blown away in the wind. However, a totally reliable eye-witness asserts that the winning ticket was a number between 4 and 10.

Joe claims he must have won and sues the organisers. In this case the prosecution hypothesis H is "Joe won the raffle" (i.e the winning ticket was 3, 4, or 5).

Joe's lawyer provides the following argument to support the claim:

> "We have two alternative hypotheses. Either Joe won the lottery (H) or Jane won the lottery (H'). We have the evidence that the winning ticket was a number between 4 and 10. Joe has numbers 4 and 5 so the evidence clearly supports H because the LR for H against H' is greater than 1. We compute the LR as follows:
>
> - Probability of the evidence given that Joe won (H) is 2/3 because if Joe won then there is a 2/3 chance the winning number was 4 or 5 (i.e. 2 of his 3 tickets could have won).
>
> - Probability of the evidence given that Jane won (H') is 1/2 because if Jane won then there is a 1/2 chance the winning number was 6.
>
> Hence the LR is 2/3 divided by ½ which is equal to 4/3"

As the LR is greater than 1 the evidence does indeed support the hypothesis that Joe won over the hypothesis that Jane won.

However, the problem with this argument is that the prosecution has chosen an alternative hypothesis that is not the negation of the prosecution hypothesis. The hypothesis "Joe did not win the raffle" is the hypothesis that the winning ticket was one of the seven numbers 1,2,6,7,8,9,10

The probability of the evidence given the winning tickets was one of the seven numbers 1,2,6,7,8,9,10 is 5/7.

So, the LR for Joe winning against Joe not winning is actually 2/3 divided by 5/7, which is equal to 14/15, a number less than 1.

By Bayes theorem the posterior odds of Joe winning (after getting the evidence) therefore actually *decreases* from a prior of 3 to 7 to a posterior of 2 to 5. Translating this into probabilities it means that the probability of Joe winning drops from 0.3 to 0.286 after getting the evidence.

So we have proven, that by choosing an alternative hypothesis H' that is not the negation of the prosecution hypothesis H it is possible to have LR>1 (meaning that the evidence provides support for H over H') even though the evidence leads to a decrease in the probability of H being true, i.e. the evidence does NOT support H. The evidence supports the negation of H irrespective of which cherry-picked alternative is considered.

**A2.2 When a LR of 1 is still probative**

We again use the lottery example with Joe having tickets 3,4 and 5, but suppose that in this case the reliable evidence is that the winning number was a number less than 7.

This time it is the defence that cherry picks an alternative hypothesis that is not the negation of H (Joe won). They find that Janet had the numbers 1, 2, 6. In this case the Defence lawyers argues:

> "We have two alternative hypotheses. Either Joe won the lottery or Janet won the lottery. We have the evidence that the winning ticket was a number less than 7. We compute the LR as follows:
>
> - Probability of the evidence given that Joe won is 1 because if Joe won then it is certain that the winning number was less than 7.
>
> - Probability of the evidence given that Jane won is 1 because if Jane won then it is certain that the winning number was less than 7.
>
> Hence the LR is 1, meaning the evidence has no probative value and therefore should not be considered."

However, again the problem with this argument is that the defence has chosen an alternative hypothesis that is not the negation of the prosecution hypothesis. The hypothesis "Joe did not win the raffle" is the hypothesis that the winning ticket was one of the seven numbers 1,2,6,7,8,9,10. The probability of the evidence (winning ticket less than 7) given the winning ticket was one of the seven numbers 1,2,6,7,8,9,10 is 3/7.

So, the LR for 'Joe winning' against 'Joe not winning' is 1 divided by 3/7 which is equal to 7/3. The evidence certainly does have probative value since the probability of Joe winning increases. By Bayes theorem the posterior odds of Joe winning (after getting the evidence) therefore actually increases from a prior of 3 to 7 to a posterior of 1 to 1 ('evens'). Translating this into probabilities it means that the probability of Joe winning increases from 0.3 to 0.5 after getting the evidence.

**Appendix 3: When the Likelihood Ratio for a suspect being a contributor to a mixed DNA profile is meaningless without strong prior information**

In what follows we create the simplest possible hypothetical example to demonstrate a fundamental limitation of the LR as a measure of probative value for mixed profile evidence when there are two unknown contributors. We consider a single locus and strip away all the usual complexities associated with the necessary calculations (such as: different peak heights, different population and allele frequencies, etc). Although the assumptions are unrealistic, they do not invalidate the thrust of the argument.

Suppose we find the following alleles of equal peak heights at one locus in a mixture:

{7,8,9,10}

Suppose we can also ignore the possibility of homozygotes at this locus and that we know for certain it is 2-person mixture.

Suppose that there are only 10 possible genotypes for this locus, namely:

(7,8), (7, 9), (7, 10), (8,9), (8, 10), (9, 10), (5,6), (6,7), (10,11), (11,12)

and that each has the same frequency (i.e. 1/10)

Then there are 3 possible combinations for the 2 contributors that could have led to the mixture, namely:

- (7,8), (9,10)
- (7,9), (8, 10)
- (7,10), (8,9)

Now suppose that there is a suspect Fred whose genotype at this locus is (7,8). This genotype is present in only the first of the 3 pairs. As we can assume each of the pairs is equally likely it follows immediately that – in the absence of any other information – that there is a probability of just 1/3 that a person with Fred's genotype is even in the mixture. This is crucial. But what is the LR for the evidence?

Let H be the hypothesis "Fred is a contributor to the mixture" and let E be the evidence "the mixture is {7,8,9,10}"

Then the LR we wish to compute is the $P(E \mid H)$ divided by $P(E \mid \text{not } H)$.

The first potential confusion is that the hypothesis H is easily confused with the slightly different prosecution hypothesis H'

> H': "A person with Fred's genotype (which may or may not be Fred) is one of the contributors to the mixture."

As we already noted, with the above assumptions – and in the absence of any other information – we can directly conclude that the posterior probability of H' after observing the evidence E is 1/3, i.e.

$$P(H' \text{ given } E) = 1/3 \qquad (1)$$

For both H and H' we want to find the LR for the hypothesis against its negation. **But crucially it turns out that, although the LRs are similar, it is only the latter which can be used meaningfully in the absence of any other evidence about Fred.**

First note that the numerator of the LR is the same in each case:

$$P(E \mid H) = P(E \mid H') = 1/10 \qquad (2)$$

This is because in each case this is the probability that the mixture profile is {7,8,9,10} given that one contributor has genotype (7,8). This is simply the probability that the other contributor has genotype (9,10). This 1/10.

However, the numerators of the respective LRs, namely P(E | not H) and P(E | not H') are not the same. We first calculate the latter because it is simpler.

**Calculating P(E | not Hp') and hence the LR for Hp'**

First note that, based on the assumption that there are 10 possible different genotypes, it follows that the number of different pairs of genotypes is:

$$10^2 - \binom{10}{2} = 100 - \frac{10 \times 9}{2 \times 1} = 55$$

Of these, 45 do not contain the genotype (7,8). Of these remaining 45 only 2 pairs, namely {(7,9),(8,10)} and {(7,10), (8,9)}, result in the mixture {7,8,9,10}. Hence

$$P(E \mid \text{not } H') = 2/45 \qquad (3)$$

It follows from (2) and (3) that

$$\text{LR for H' against not H'} = \frac{P(E \mid H')}{P(E \mid \text{not } H')} = \frac{1/10}{2/45} = \frac{9}{4}$$

So, as expected, the LR supports the prosecution hypothesis because it is greater than 1.

But does the LR in this case tell us anything about the probability of the prosecution hypothesis H'? It turns out it does in this case **because, in the absence of any other evidence against Fred, we know what the 'correct' prior probability for H' is**. As explained above there are 55 different possible genotype pairs in a 2-person mixture, of which 10 involve the genotype (7,8). So, the prior probability P(H'), i.e. the prior probability that the genotype (7,8) is one of the pairs in a mixture, is simply 10/55 = 2/11. This is the same as odds 2 to 9. Using the LR we can compute the posterior odds as

$$\text{posterior odds} = \text{LR} \times \text{prior odds} = \frac{9}{4} \times \frac{2}{9} = \frac{1}{2}$$

In other words, the posterior probability is 1/3. But – and this is the crucial point – this is exactly the probability that we already determined directly in (1).

This validates the LR calculation.

But it also confirms that, despite an LR of greater than 1, the prosecution hypothesis H' is still unlikely to be true given the evidence.

**Calculating P(E | not H) and the LR for H**

P(E |not H) is the probability that the mixture profile is {7,8,9,10} given that neither contributor is Fred. There are three ways in which we can arrive at this mixture profile if Fred is not a contributor:

1. The contributors have genotypes (7,9) (8,10)
2. The contributors have genotypes (7,10) (8,9)
3. The contributors have genotypes (7,8) (9,10) but the (7,8) person is NOT Fred.

Each of the first two has probability 1/45 as explained in the H' case. The third has probability 1/10 times 1/45 because there is a probability of 1/10 that the genotype will be (7,8) in a person who is not Fred.

Hence:
$$P(E\,|not\,H) = \frac{2}{45} + \frac{1}{450} = \frac{21}{450} \qquad (4)$$

It now follows from (2) and (4) that

$$\text{LR for H against not H} = \frac{P(E\,|H)}{P(E|not\,H)} = \frac{1/10}{21/450} = \frac{45}{21}$$

This LR is slightly lower than in the case of H'. However, it **is completely meaningless in the absence of other evidence against Fred**. This is because, unlike for H' we cannot consider any normative prior for H without knowing the number of other potential contributors. Only by considering such numbers can we 'validate' the LR.

For example, we could assume that there are 1000 possible contributors (including Fred) to the mixture. Then there are is a prior probability of 1/500 that Fred is a contributor to a 2-person mixture. Combining this with the LR we get a posterior probability for H of approximately 1/232. So, despite the LR supporting H, it is very unlikely Fred is a contributor.

**Conclusion**

Without strong prior assumptions about the probability that a suspect <u>is a contributor</u> to a 2-person mixture the LR for the hypothesis that the suspect is a contributor against its negation is essentially meaningless. In contrast, the LR for the hypothesis that the suspect's profile is contained in the mixture is meaningful without any prior information. But in this case, despite a LR>1 , the posterior probability will be low.

While in the example we considered only a single locus 'match' – meaning the LT was not especially large, in the next appendix we show that in theory, as the number of loci in the mixture increase (and the suspect's alleles are included in each loci of the mixture) the LR will become enormous, even though posterior probability of H' in such cases will also be very low in the absence of other evidence.

# Appendix 4: A very high LR 'supporting inclusion' but a very low posterior probability of inclusion

*The objective of this example is to demonstrate – using a simple analogy with numbered balls - the limitations of what can be claimed regarding a high likelihood ratio (LR) for a mixture DNA profile..*

We use the same analogy of numbered balls as in the last example of Appendix 1, as this will make clear how different the mixture profile 'inclusion' scenario case is from the single profile 'match' case.

We start with a simple case of three balls and then generalise it.

**Simple 3-ball case**

Suppose we know that every person is given three balls labelled ball1, ball2, ball3 and that each ball contains a random number from 0 to 9 (each being equally likely).

Suppose Fred's balls respectively have numbers 1, 2 and 3. Then we say the sequence (1 2 3) is "Fred's number profile".

There are $10^3$ =1000 different possible number profiles (0 0 0), (0 0 1), (0 0 2), …, (9 9 9)

Now suppose that we know two people (who we call 'contributors' one of whom may or may not be Fred) have thrown their balls into labelled pots, i.e. two people throw their ball1 into pot 1, and their ball2 into pot 2 and their ball3 into pot 3. At the end of this we see that the pots contain a mixed profile of the form:

| Pot 1 | Pot 2 | Pot 3 |
|---|---|---|
| 1, 7 | 2, 8 | 3, 9 |

The first thing we note about this mixed profile is that Fred's numbers are INCLUDED in each pot, so it is possible that he is one of the two contributors. But what does the evidence tell us about the hypothesis

> H: "A person with Fred's number profile (who may or may not be Fred) is one of the contributors to the mixture."

In this case there are 4 different profile pairs that would result in this mixture, namely

- (1 2 3), (7 8 9)
- (1 2 9), (7 8 3)
- (1 8 3), (7 2 9)
- (1 8 9), (7 2 3)

So, although Fred's profile (1 2 3) has matching numbers for each pot of the mixture, there is only a ¼ probability that somebody with that profile is actually in the mixture (it is also important to note that, if the profile (1 2 3) is in the mixture then there is also no certainty that this comes from Fred; it could come from anybody with the same profile as the Fred).

What we have demonstrated is that the posterior probability of H given the evidence E (of the particular mixture profile) is ¼.

To determine the prior probability H (before any evidence) we have to calculate the probability that a suspect's profile will be in an unknown 2-person mixture. We note that the number of different possible 2-person mixture profiles is simply the number of combinations of the 1000 possible profiles 2 at a time. This is defined as:

$$\binom{1000}{2} = \frac{1000 \times 999}{2 \times 1}$$

Of these, the Fred's profile appears in exactly 1000 as it can be paired with every possible profile (including the same one as the Fred's).

Hence the prior P(H) is

$$P(H) = \frac{1000}{\binom{1000}{2}} = \frac{1000 \times 2}{1000 \times 999} = \frac{2}{999}$$

This means that the probability of 'not H' is

$$P(\overline{H}) = \frac{997}{999}$$

So the prior 'odds' of Fred's profile being in a 2-person mixture are 2 to 997

Now let us calculate the LR for H against not H with the evidence E (a 2-person mixture as in the above table). Clearly

$$P(E|H) = \frac{1}{1000}$$

because, if the defendant's profile (1 2 3) is in the mixture, then exactly one of the 1000 possible profiles for the second contributor, namely (7 8 9) in the case, must also be in the mixture.

But what is $P(E|\overline{H})$?

As there are $\binom{1000}{2}$ total profile pairs and 1000 of these contain Fred's it follows that there are

$$\binom{1000}{2} - 1000$$

profile pairs that do not contain Fred's. Of these there are exactly 3 that result in the evidence E, namely

- (1 2 9), (7 8 3)
- (1 8 3), (7 2 9)

- (1 8 9), (7 2 3)

Hence:

$$P(E|\bar{H}) = \frac{3}{\binom{1000}{2} - 1000} = \frac{3}{\frac{1000 \times 999}{2} - 1000} = \frac{3}{500 \times 997}$$

Hence, the LR in this case is:

$$\frac{P(E|H)}{P(E|\bar{H})} = \frac{\frac{1}{1000}}{\frac{3}{500 \times 997}} = \frac{997}{6} = 166.17$$

Now we already know that the posterior probability of H given E is ¼ because the suspect's profile is in exactly one of the four possible pairs of profiles that result in the mixture. But we can now also confirm this with Bayes' theorem because the prior odds are 2 to 997 and when we multiply by the LR of 997/6 we get posterior odds 1 to 3 for H against not H , i.e posterior probability for H of ¼ .

*Conclusion*: In the case of a 3-ball profile if the defendant's balls are contained in each pot of a 2-person mixture the LR for the hypothesis that the defendant's profile is in the mixture (against it not being in the mixture) is high (166.17) but the probability that his profile is actually in the mixture is still only ¼. Moreover any person with the same number profile as Fred is equally likely to have been a contributor.

**Generalising to an arbitrary number of balls**

We can generalise this to any number of balls. Suppose in the case of 10 ball profile Fred's, profile is (1 2 3 1 2 3 0 7 8 9) and that his balls appear in every pot of a 2-person profile:

| P1 | P2 | P3 | P4 | P5 | P6 | P7 | P8 | P9 | P10 |
|---|---|---|---|---|---|---|---|---|---|
| 1, 7 | 2, 8 | 3, 9 | 1, 5 | 2, 6 | 3, 0 | 0, 9 | 7, 2 | 8, 5 | 9, 1 |

Assuming that two different numbers appear in each pot there are $2^9$ different profile pairs that result in the same mixture[1]. This is because once the first position is fixed the other 9 positions can be filled by any permutation of the two available numbers in that position. So, in general, if we have n-ball profiles and if there are different numbers in each pot of the mixture then there are $2^{n-1}$ different profile pairs that could result in the same mixture. Hence, although the suspect's profile has matching numbers for each of the n pots of the mixture, there is only a very small probability $\frac{1}{2^{n-1}}$ that somebody with that profile is actually in the mixture.

Yet the LR is incredibly high. Before we calculate it we note that the prior for H is

$$P(H) = \frac{10^n}{\binom{10^n}{2}} = \frac{10^n \times 2}{10^n \times (10^n - 1)} = \frac{2}{10^n - 1}$$

because of the $\binom{10^n}{2}$ possible pairs of profiles, the defendant's profile appears in exactly $10^n$ as it can be paired with every possible profile (including the same one as the defendant's).

---

[1] If, there are k pots that contain the same number, then there are $2^{9-k}$ different profile pairs that result in the same mixture; for example if PI contains 1,1 and P2 contains 2,2 then there are $2^7$ different profile pairs that result in the same mixture

This also means that

$$P(\bar{H}) = \frac{10^n - 3}{10^n - 1}$$

And so the prior 'odds' of H to not H are 2 to $(10^n - 3)$

Clearly

$$P(E|H) = \frac{1}{10^n}$$

because, if the defendant's profile is in the mixture, then exactly one of the $10^n$ possible profiles for the second contributor must also be in the mixture (in the example given the one other contributor profile would have to be (7 8 9 5 6 0 9 2 5 1) since the Fred's profile is (1 2 3 1 2 3 0 7 8 9)).

But what is $P(E|\bar{H})$?

As there are $\binom{10^n}{2}$ total profile pairs and $10^n$ of these contain the suspect's profile it follows that there are

$$\binom{10^n}{2} - 10^n$$

profile pairs that do not contain the suspect's. Now we know that there are $2^{n-1}$ different profile pairs that would result in the evidence E, and exactly one of these contains the suspect's profile. Hence, there are exactly $2^{n-1} - 1$ profile pairs that result in the evidence E that do not contain the suspect's profile.

Hence:

$$P(E|\bar{H}) = \frac{2^{n-1} - 1}{\binom{10^n}{2} - 10^n} = \frac{2^{n-1} - 1}{\frac{10^n \times (10^n - 1)}{2} - 10^n} = \frac{2^{n-1} - 1}{\frac{10^n}{2} \times (10^n - 3)}$$

The LR is thus

$$\frac{P(E|H)}{P(E|\bar{H})} = \frac{1}{10^n} \times \frac{\frac{10^n}{2} \times (10^n - 3)}{2^{n-1} - 1} = \frac{10^n - 3}{2(2^{n-1} - 1)}$$

So, if n=24 the LR is approximately $6 \times 10^{16}$ - an astronomically high number. But, we already know that the posterior probability of H given E is incredibly low, namely $\frac{1}{2^{n-1}}$ (which for n=24 is approx one in 8.4 million) because the suspect's profile is in exactly one of the $2^{n-1}$ possible pairs of profiles that result in the mixture. But we can now also confirm this with Bayes' theorem because the prior odds of H to not H are 2 to $(10^n - 3)$ and when we multiply by the LR above we get posterior odds 1 to $2^{n-1} - 1$, i.e. the posterior probability of H is $\frac{1}{2^{n-1}}$.

Before discussing some of the simplified assumptions it is fair to say this is a close analogy to the case of a 2-person mixed DNA profile[2]. We can think of the number on any of Fred's balls (*i*) as corresponding to his allele combination at locus *i* – so instead of a single digit number this will be a pair like (15,17) or (17,17) and typically at any locus any given allele combination will be shared with between 5% and 20% of the population – so not much different to the 1 in 10 'match' probability we are assuming here.

So, if we are sure we have a 2-person DNA mixed profile and if we are sure that Fred's alleles are included on all 24 loci then we can conclude that there is a very high LR in favour of the hypothesis "Fred plus one unknown contributor" over the hypothesis "Two unknown contributors". ***But we can also conclude there is a very low probability that either of the contributors has a DNA profile that matches that of Fred***. What does this mean for the DNA evidence?

The defence could reasonably argue that the high LR is irrelevant and that what really matters is that there is a very low posterior probability that *either* contributor's DNA profile is a match to Fred's.

The prosecution could argue that, even in the absence of any other evidence the prior and posterior probability of Fred being a contributor should take account of the number of people for whom it was physically possible to have been at the crime scene. If, for example, this number was proven to be at most 1000 including the suspect, then the maximum number of different possible profiles in the mixture would be 1000 and not $10^{24}$ in the case where the number of loci (balls) in the profile n=24. In this case they can argue that the *posterior* probability that Fred is a contributor is very high because the LR is $6 \times 10^{16}$.

**The simplifications**:
- The most obvious simplification (as already noted) is that the calculations assumes a *different* numbered ball to Fred's is in each pot. In principle two people could of course have the same number on any pot (as two different people could have either the same allele combination or a common allele in a pair). So, for example, pot 1 could contain 2 balls with the number 1 etc. In this case the number of possible different profile pairs that could have led to the mixture is $2^{n-k-1}$ (and not $2^{n-1}$) assuming that there are k positions in which the numbers are the same. This results in a higher LR.

---

[2] This does NOT apply to the case where there is one known (major) contributor, but it does apply to the case where there is one known (major) contributor and two unknown (minor) contributor assuming we ignore the major contributor alleles.

- We are assuming binary decisions about presence/absence of numbers. For probabilistic genotyping systems the (numeric value) peak heights for alleles are used in the overall probability calculations.

Neither of these simplifications changes the general thrust of the example. It will still be the case that, for a 24-loci mixture, the inclusion of Fred's allele's at each loci will lead to a very high LR but also a very low probability that one of the two contributors has a profile matching Fred's.

Other simplifications made are actually favourable to the prosecution hypothesis. Specifically:

- We are assuming exhaustive hypotheses H and not H which enables us to make conclusions about the posterior probability of H from the LR. In practice, as we have noted, the defence hypothesis ('two unknown contributors') is not the negation of the prosecution hypothesis due to the possibility of a) people related to the suspect being in the mixture; and b) more than two people being in the mixture.
- We are assuming a 2-person mixture. When there are 3 or more the calculations become much complex, but are generally much more favourable to the defence hypothesis. This is because the more contributors there are the more likely it is for any profile to be included in the mixture. Consider the extreme case of a 10-person mixture where in each pot the numbers are {0, 1, 2, 3, 4, 5, 6, 7, 8, 9}. Then every possible profile is potentially included in the mixture.

**Appendix 5: Multiple different prosecution propositions for the same evidence – each with very high LR even though at most one proposition is true and the others must be false**

Using the same example as Appendix 4 where each person has a sequence of 20 balls. To make the analogy closer to the DNA analogy we will assume that each ball i (representing a locus) is labelled with one of 10 letters $A_i, B_i, C_i, D_i, E_i, F_i, G_i, H_i, I_i, J_i$ and that each is equally likely (so these 'correspond' to genotypes with assumed frequency 1/10).

Suppose one or more people throw all their balls into pots P1, P2, …P20 where pot Pi contains ball i.

Now suppose that a single ball is taken from each pot with the following result:

| P1 | P2 | P3 | P4 | P5 | P6 | P7 | P8 | P9 | P10 | P11 | P12 | P13 | P14 | P15 | P16 | P17 | P18 | P19 | P20 |
|---|---|---|---|---|---|---|---|---|---|---|---|---|---|---|---|---|---|---|---|
| $A_1$ | $E_2$ | $A_3$ | $D_4$ | $J_5$ | $D_6$ | $C_7$ | $G_8$ | $A_9$ | $E_{10}$ | $B_{11}$ | $D_{12}$ | $J_{13}$ | $D1_{14}$ | $J_{15}$ | $D_{16}$ | $C_{17}$ | $B_{18}$ | $B_{19}$ | $E_{20}$ |

So our evidence E is that the above sequence of balls is found in the respective pots.

A search finds that Fred's ball sequence is an exact match of the balls selected, namely:

| $A_1$ | $E_2$ | $A_3$ | $D_4$ | $J_5$ | $D_6$ | $C_7$ | $G_8$ | $A_9$ | $E_{10}$ | $B_{11}$ | $D_{12}$ | $J_{13}$ | $D1_{14}$ | $J_{15}$ | $D_{16}$ | $C_{17}$ | $B_{18}$ | $B_{19}$ | $E_{20}$ |
|---|---|---|---|---|---|---|---|---|---|---|---|---|---|---|---|---|---|---|---|

So: Is Fred a contributor?

The problem is we do not know the exact number of contributors.

But we can calculate the LR based on different numbers k of contributors as follows

| k | H1 | H2 | P(E\|H1) | P(E\|H2) | LR |
|---|---|---|---|---|---|
| 1 | Fred is contributor | Unknown contributor | 1 | $10^{-20}$ | **$10^{20}$** |
| 2 | Fred + unknown contributor | 2 unknown contributors | $6.42*10^{-6}$ | $10^{-20}$ | **$6.42*10^{14}$** |
| 3 | Fred + 2 unknown contributors | 3 unknown contributors | $1.10*10^{-8}$ | $10^{-20}$ | **$1.10*10^{12}$** |
| 4 | Fred + 3 unknown contributors | 4 unknown contributors | $1.73*10^{-10}$ | $10^{-20}$ | **$1.73*10^{10}$** |
| 5 | Fred + 4 unknown contributors | 5 unknown contributors | $8.7*10^{-12}$ | $10^{-20}$ | **$8.77*10^{8}$** |
| 6 | Fred + 5 unknown contributors | 6 unknown contributors | $9.09*10^{-13}$ | $10^{-20}$ | **$9.09*10^{7}$** |
| 7 | Fred + 6 unknown contributors | 7 unknown contributors | $1.52E*10^{-13}$ | $10^{-20}$ | **$1.5*10^{7}$** |
| 8 | Fred + 7 unknown contributors | 8 unknown contributors | $3.53*10^{-14}$ | $10^{-20}$ | **$3.53*10^{6}$** |
| 9 | Fred + 8 unknown contributors | 9 unknown contributors | $1.05*10^{-14}$ | $10^{-20}$ | **$1.05*10^{6}$** |
| 10 | Fred + 9 unknown contributors | 10 unknown contributors | $3.76*10^{-15}$ | $10^{-20}$ | **$3.76*10^{5}$** |
| 15 | Fred + 14 unknown contributors | 15 unknown contributors | $1.21*10^{-16}$ | $10^{-20}$ | **$1.21E10^{4}$** |
| 20 | Fred + 19 unknown contributors | 20 unknown contributors | $4.69*10^{-18}$ | $10^{-20}$ | **469** |
| 25 | Fred + 24 unknown contributors | 25 unknown contributors | $1.90*10^{-18}$ | $10^{-20}$ | **190** |
| 40 | Fred + 39 unknown contributors | 40 unknown contributors | $5.79*10^{-19}$ | $10^{-20}$ | **58** |
| 80 | Fred + 79 unknown contributors | 80 unknown contributors | $8.43*10^{-20}$ | $10^{-20}$ | **8** |

And this exposes the problem with the meaning of high LRs. Let us, for example, consider just two of the possibilities:

By assuming 3 contributors the prosecution can claim:

> "The evidence is $1.1*10^{12}$ (more than a trillion) times more likely if Fred is a contributor than if he is not."

By assuming 4 contributors the prosecution can claim:

> "The evidence is $1.73*10^{10}$ (more than 10 billion) times more likely if Fred is a contributor than if he is not."

But at least one of the hypotheses: "Fred + 2 unknowns" or "Fred + 3 unknowns" is false. If, say, the first is false then it follows that:

> Despite the fact that the evidence is more than a trillion times more likely if a Fred is a contributor than if he is not, ***it is CERTAIN than Fred is not actually a contributor***.

So we have multiple different prosecution hypotheses (at most one of which can be true) all yielding incredibly high LRs.

**The maths behind it**

In what follows we assume for each k (total number contributors):

H1: "Fred + (k-1) unknowns are contributors"

H2: "k Unknowns are contributor"

*When k=1*

> P(E | H1) = 1 because if Fred is THE contributor then his sequence of balls is the only possible sequence that can be selected.
>
> P(E | H2) = $(1/10)^{20}$ because if some unknown person is a contributor there is a 1/10 probability that the number would be the same as Fred's for that pot.
>
> Hence assuming exactly one contributor LR = $10^{-20}$

*When k=2*

> To calculate P(E|H1) think about a single pot, say pot 1. We know that the letter A must be one of the 2 balls there because that is Fred's number in pot 1. There is a 1/10 probability the other numbered ball is A. There is probability 1/2 of selecting the ball known to be A and a ½ probability of selecting the other ball, so there is a ½ + ½(1/10) probability that the selected ball from pot 1 is numbered A. This is 0.55. As there are 20 pots:
>
> P(E | H1) = $(0.55)^{20}$
>
> But P(E| H2) is the same as in the 1 person case, namely $(1/10)^{20}$ because if two unknowns are contributors each has a 1/10 probability of being the same as Fred's for that pot and each has a ½ probability of being selected.

Hence LR=$6.42 \times 10^{14}$

General case k

To calculate P(E | H1) think about a single pot, say pot 1. We know that the letter A must be one of the k balls there because that is Fred's number in pot 1. Each of the other (k-1) balls has a 1/10 probability of being labelled A. There is probability 1/k of selecting the ball known to be A and a 1/k probability of selecting each of the other (k-1) balls, so there is a 1/k + (k-1)*(1/k*1/10) probability that the selected ball from pot 1 is A. Hence

P(E | H1) = (1/k + (k-1)*(1/k*1/10))^20

But P(E | H2) is the same irrespective of k, namely $10^{-20}$

Hence LR = [ (1/k + (k-1)*(1/k*1/10))^20 ] / $10^{-20}$